\documentclass[12pt]{article}
\usepackage{amsmath,amssymb,amsthm,graphicx,dsfont,hyperref,amsfonts}
\usepackage[mathscr]{eucal}
\usepackage{color}

\numberwithin{equation}{section}
\theoremstyle{plain} 
\theoremstyle{definition}
\newtheorem{thm}{Theorem}[section]

\newtheorem{prop}[thm]{Proposition}

\theoremstyle{definition}

\theoremstyle{remark}

\newcommand{\appendices}{
  \renewcommand{\thesection}{\Alph{section}}
  \setcounter{section}{0}
  \newpage
}

\newcommand{\five}[1]{\ensuremath{\hspace{-3pt}\left.^{(5)}\hspace{-1pt}#1\right.}}
\newcommand{\four}[1]{\ensuremath{\hspace{-3pt}\left.^{(4)}\hspace{-1pt}#1\right.}}

\newcommand{\C}{\mathbb{C}}
\newcommand{\CP}{\mathbb{CP}}

\newcommand{\Z}{\mathbb{Z}}

\renewcommand{\O}{\mathcal{O}}
\newcommand{\pp}[2]{\frac{\partial #1}{\partial #2}}

\DeclareMathOperator{\Tr}{Tr}

\newcommand{\be}{\begin{equation}}
\newcommand{\ee}{\end{equation}}
\newcommand{\bal}{\begin{align}}
\newcommand{\eal}{\end{align}}
\newcommand{\ben}{\begin{equation*}}
\newcommand{\een}{\end{equation*}}
\newcommand{\bea}{\begin{eqnarray}}
\newcommand{\eea}{\end{eqnarray}}
\newcommand{\bean}{\begin{eqnarray*}}
\newcommand{\eean}{\end{eqnarray*}}
\newcommand{\bes}{\begin{subequations}}
\newcommand{\ees}{\end{subequations}}

\definecolor{newtext}{rgb}{0.85,0,0}
\definecolor{txt}{rgb}{0,0,0.85}
\newcommand{\newtext}[1]{{\color{newtext} #1}}

\begin{document}

\begin{titlepage}
\begin{flushright}
IPMU 10-0213
\end{flushright}
\bigskip
\bigskip\bigskip\bigskip
\centerline{\Large Can you hear the shape of dual geometries?}
\bigskip\bigskip\bigskip
\bigskip\bigskip\bigskip

\centerline{{\bf Richard Eager${}^{1,2}$, Michael Gary${}^{1}$}, and {\bf Matthew M. Roberts${}^{3}$}}
\medskip
\centerline{\em${}^{1}$ Department of Physics}
\centerline{\em University of California}
\centerline{\em Santa Barbara, CA 93106-4030, USA}

\bigskip

\centerline{\em${}^2$ Institute for the Physics and Mathematics of the Universe}
\centerline{\em University of Tokyo}
\centerline{\em Kashiwa, 277-8583, Japan}
\bigskip

\centerline{\em${}^3$ Department of Physics}
\centerline{\em New York University}
\centerline{\em 4 Washington Place}
\centerline{\em New York, NY 10003, USA}
\bigskip
\centerline{\tt reager@physics.ucsb.edu,~mgary@physics.ucsb.edu,}
\centerline{\tt matthew.roberts@nyu.edu}

\begin{abstract}
We compute the sub-leading terms in the Tian-Yau-Zelditch asymptotic expansion of the partition function for dual giant gravitons on $AdS_5 \times L^5$ and provide a bulk interpretation in terms of curvature invariants. We accomplish this by relating the partition function of dual giant gravitons to the Hilbert series for mesonic operators in the CFT. The coefficients of the subleading terms encode integrated curvature invariants of $L^5.$  In the same spirit of Martelli, Sparks and Yau, we are able to compute these integrated curvature invariants without explicit knowledge of the Sasaki-Einstein metric on $L^5$. These curvature invariants contribute to the $1/N^2$ corrections of the difference of the $4D$ anomaly coefficients $a$ and $c$ recently found by Liu and Minasian, which we now have a purely field theoretic method of calculating.  
\end{abstract}
\end{titlepage}
\section{Introduction}
Following the discovery of the AdS/CFT correspondence on $AdS_5\times S^5$ \cite{Maldacena:1997re}, the duality was generalized to compactifications of the form $AdS_5\times L^5$, where $L^5$ is an arbitrary Sasaki-Einstein manifold \cite{Morrison:1998cs,Acharya:1998db}.  The Sasaki-Einstein condition ensures that the low energy supergravity theory on $L^5$ preserves 8 supercharges.  This supersymmetry condition has a geometric interpretation, that the cone $X^6$ over $L^5$ is a Calabi-Yau singularity.  The dual field theory is obtained from the low-energy field theory on the world-volume of $N$ D3-branes placed at the Calabi-Yau singularity. By studying fluctuations on the string theory side one finds that information about the $L^5$ factor must be encoded in the particle content and spectrum for the dual CFT. This motivates one to ask, paralleling the iconic question in spectral theory \cite{Kac:1966xd}, ``is it possible to hear the shape of a dual geometry?'' 

In fact, the analogy to the classic question of Kac in spectral theory is remarkably close, as we make use of the Hilbert series, a holomorphic analog of the heat kernel. Generalizing the works of Martelli, Sparks and Yau \cite{Martelli:2005tp,Martelli:2006yb,Martelli:2006vh}, we relate curvature invariants of $L^5$ to the number of mesonic operators in the CFT by counting the asymptotic number of holomorphic functions on the Calabi-Yau cone over $L^5$. By computing sub-leading $1/N^2$ corrections to the asymptotic number of long mesonic operators, we are able to compute $\int_{L^5}\mathrm{Riem}^2$ from purely CFT data without reference to an explicit metric on $L^5$.

The AdS/CFT correspondence predicts that the volume of the horizon manifold is inversely relate to the $a$-central charge of the dual gauge theory
$$Vol(L^5) = \frac{\pi^3 N^2}{4 a}.$$
Recently this piece of the AdS/CFT conjecture was mathematically proven \cite{Eager:2010yu} by analyzing the leading order behavior of the Hilbert series.  The sub-leading coefficients of the Hilbert series contain a wealth of geometric information as well. The main result of this paper is the following relation between the coefficients of the singular terms in the Hilbert series of the quiver gauge theory and curvature invariants of $L^5$
\begin{equation}
 \Tr_{\mathcal{H}}e^{-\beta \hat{H}} \propto \lim_{s\rightarrow0}\left(\frac{2\pi}{3}\right)^3 H_{0,0}(Q,e^{-s}) \nonumber\ee
\be=\frac{\mathrm{vol}(L^5)}{s^3}+\frac{\mathrm{vol}(L^5)}{s^2}+\left(\frac{91}{216}\mathrm{vol}(L^5)+\frac{1}{1728}\int_{L^5}\mathrm{Riem}^2(L^5)\right)\frac{1}{s}+\mathcal{O}(1) \label{thmconj}
\end{equation}
This is also an asymptotic expansion of the quantum partition function for dual giant gravitons.
We are able to prove this when $L^5$ is a regular Sasaki-Einstein manifold (that is when it is the total space of a regular $U(1)$ bundle over a 4d K\"ahler-Einstein base $B^4$). Our proof uses the Tian-Yau-Zelditch (TYZ) expansion of the heat kernel \cite{Tian,Zelditch,Catlin,Lu} which relates the Hilbert series to the curvatures of the 4d base and then we ``undo'' the Kaluza-Klein reduction to rewrite this in terms of the 5d total space.  For several irregular Sasaki-Einstein manifolds where an explicit metric is known \cite{Gauntlett:2004yd} we have verified by direct calculation that (\ref{thmconj}) also holds. We therefore conjecture that (\ref{thmconj}) holds not only for $L^5$ regular Sasaki-Einstein, but for quasi-regular and irregular Sasaki-Einstein manifolds as well.

The physics motivating our proof is as follows. First, we believe from AdS/CFT that the partition functions for dual giant gravitons is the Hilbert series for the quiver gauge theory. Indeed, Martelli and Sparks showed that BPS dual giant gravitons act as point particles in $\mathrm{time}\times L^5$ \cite{Martelli:2006vh}. It is clear that a point particle moving on $L^5$ can be Kaluza-Klein reduced to an electrically charged point particle moving in an electromagnetic field in $B^4$. This scenario was analyzed by Douglas and Klevtsov \cite{Douglas:2008pz} and they re-derived the TYZ expansion of the partition function for a particle in a magnetic field.  Therefore the problem of counting mesonic operators in the CFT is reduced to counting the number of states in the lowest Landau level in the base space $B^4$. Douglas and Klevtsov relate the expansion of the Landau level counting to curvature invariants on the base space $B^4$. However, the requirement that $L^5$ be Sasaki-Einstein and that $B^4$ be at least locally K\"{a}hler-Einstein with a $U(1)$ line bundle is a strong enough condition that it fixes a relationship between 4d and 5d curvature invariants. Since in both the regular and irregular cases, $L^5$ is smooth, we conjecture that the expansion written in 5d curvatures also holds in the irregular case.

The outline of the rest of the paper is as follows.
In section \ref{HilbertSeries} we review the Hilbert series and the correspondence between mesonic operators and holomorphic functions on the Calabi-Yau cone. Section \ref{BPS} reviews the results of Martelli and Sparks \cite{Martelli:2006vh}, which relates the Hilbert series to the partition function of dual giant gravitons. We then connect their partition function to that of the integer quantum hall system studied by Douglas and Klevtsov \cite{Douglas:2008pz}. In section \ref{Conjecture} we express the sub-leading terms in the asymptotic expansion of the Hilbert series as geometric invariants, provide a proof of the conjectured form for regular Sasaki-Einstein manifolds, and provide evidence that the result holds more generally for quasi-regular and irregular Sasaki-Einstein manifolds as well. Finally, we conclude in section \ref{Conclusions} with more general remarks and avenues for future research. We also include an appendix which provides further details omitted from the main body of the paper. 

As this work was nearing publication, we became aware of the work of Liu and Minasian \cite{Liu:2010gz} which computes contributions to $1/N^2$ corrections to the anomaly coefficients $a$ and $c$ of the dual field theory. Such corrections correspond to higher curvature terms in the AdS gravitational action, which reduce to integrated curvature invariants in the expansion, an interpretation complementary to the work presented in this paper. Our work provides a method of calculating these curvature invariants from purely field theoretic data.

\section{The Hilbert series}\label{HilbertSeries}
In this section we consider $\mathcal{N}=1$ supersymmetric conformal field theories (SCFTs) obtained from a stack of $N$ D3-branes at a Calabi-Yau singularity $X^{6}$.  For toric Calabi-Yau 3-folds, mesonic operators in the gauge theory are in one-to-one correspondence with holomorphic functions on the Calabi-Yau manifold \cite{Hanany:2006nm,Benvenuti:2006qr,Martelli:2006vh}.  For general $\mathcal{N}=1$ superconformal quiver gauge theories we can identify mesonic operators based at a $U(N)$ gauge group with holomorphic functions on the Calabi-Yau manifold \cite{Eager:2010yu}.

We consider local Calabi-Yau singularities $X^{6} = \mathrm{Spec} R$ where $R$ is a Gorenstein ring of dimension 3.  A broad class of such singularities can be described by quiver gauge theories with superpotential algebras $A$ of the form
\begin{equation}
A=\mathrm{End}_R(R+M_1+\cdots +M_n)
\end{equation}
where $M_i$ are $R$-modules corresponding to fractional branes.  Since $\mathrm{End}_R(R)=R$, it is possible to identify closed loops based at the node corresponding to $R$ with the elements of the ring $R$.  These loops are the holomorphic functions on the Calabi-Yau singularity $X^6$, and thus the based mesonic operators in the CFT are in one-to-one correspondence with holomorphic functions on the Calabi-Yau 3-fold. 

The Hilbert series of a quiver gauge theory is defined to be
$$H(t) = \sum_{\mathcal{O} \in \text{ Mesons}} t^{R(\mathcal{O})}$$
where the sum extends over all mesonic operators $\mathcal{O}$ based at a $U(N)$ gauge group up to F-term equivalence.  For a superconformal quiver gauge theory, we define an adjacency matrix graded by R-charge,
\begin{equation}
M(Q;t)_{ij}=\sum_{e\in\mathrm{Arrows}(i\rightarrow j)}t^{R(e)}\ ,
\end{equation}
where $R(e)$ is a trial R-charge assignment for the edge $e$.  A trial R-charge is an R-charge assignment such that the quiver gauge theory has vanishing beta functions and is hence superconformal. The true R-charge assignment in the infrared is determined by the A-maximization procedure of Intriligator and Wecht \cite{Intriligator:2003jj}.  The Hilbert series of the superconformal quiver gauge theory is the $(0,0)$ component of the matrix \cite{Eager:2010yu}
\begin{equation}
H(Q;t)=\frac{1}{1-M_Q(t)+t^2 M_Q^T(t^{-1})-t^2} \qquad\ .
\end{equation}

We now explain the Hilbert series in the context of singularities $X$ that are the total space of a line bundle $\mathcal{L} \rightarrow B^4$, which is the case of primary interest throughout the remainder of this paper
\footnote{Not all Sasaki-Einstein manifolds are of this form.  The base $B^4 = \mathbb{F}_1$, known as the first del Pezzo surface, has no K\"ahler-Einstein metric.  The existence of a K\"ahler-Einstein metric in this case is obstructed by the Matsushima and Futaki theorems \cite{MR0094478,Futaki:2008vr,Gauntlett:2006vf}.  However the ``link'' of the total space of the anti-canonical line bundle over $\mathbb{F}_1$, which is called $Y^{2,1}$, does admit a Sasaki-Einstein metric \cite{Martelli:2005wy}.
}.
In this case, the corresponding Sasaki-Einstein manifold is regular. For such Calabi-Yau manifolds, the Hilbert series takes the form
\begin{equation}
\label{HilbertSum}
H(t,X) = \sum_{k = 0}^{\infty} \mathrm{dim} \; \mathcal{H}^0 (B, \mathcal{L}^{\otimes k}) t^k
\end{equation}

in this case, the TYZ asymptotic expansion of the Hilbert series is sensitive to the volume and curvature of the base $B^4$ \cite{Tian,Zelditch,Catlin,Liu:2010gz}
\begin{equation}\label{DouglasTYZ}
H(t,X) = g(k)=\frac{1}{2} \left(k^2+k\frac{R}{2}+\left(\frac{1}{24}\mathrm{Riem}^2-\frac{1}{6}\mathrm{Ric}^2  +\frac{1}{8}R^2\right)\right)+\mathcal{O}(\frac{1}{k})\ .
\end{equation}

It is also useful to consider the restriction of $\C$ to $U(1)$, which naturally gives $L^5$ as a circle bundle over $B^4$
\begin{equation}\begin{array}{ccc}
S^1 &\rightarrow& L^{5} \\
&&\downarrow\\
&&B^4\ .\end{array}
\end{equation}

 As an explicit example, consider the case $X = \C^3$, in which case $L = S^5$ and $B = \CP^2$ and the relevant line bundle is $\mathcal{L} = \O(1)$. 
\begin{equation}
\mathrm{dim} \; \mathcal{H}^0 (B, \mathcal{L}^{\otimes k}) = \frac{(k+1)(k+2)}{2}\ .
\end{equation}
Computing the Hilbert series for $\C^3$, we find
\begin{equation}
H(t,X) = \sum_{k=0}^\infty \frac{(k+1)(k+2)}{2}t^k = \frac{1}{(1-t)^3}\ .
\end{equation}

As shown by Martelli, Sparks, and Yau \cite{Martelli:2005tp,Martelli:2006yb} and will be further elaborated in section \ref{Conjecture}, the TYZ expansion determines $\mathrm{dim} \; \mathcal{H}^0(B,\mathcal{L}^{\otimes k})$ as a quadratic polynomial $ak^2+bk+c$ whose leading coefficient $a$ is proportional to the volume of the horizon manifold. Similarly, the linear term $b$ is fixed by the Calabi-Yau condition, and as we will show, the constant term $c$ is determined by integrating a linear combination of quadratic curvatures over the horizon manifold $L$. 

Given the Hilbert series $H(t)$ for a Calabi-Yau 3-algebra, which has a pole at $t=1$, the coefficients $a$, $b$, and $c$ are easily computed from the singular part of $H(e^{-s})$ near $s=0$,
\begin{equation}\label{Smalls}
\lim_{s\rightarrow0}H(e^{-s})=\frac{2a}{s^3}+\frac{b}{s^2}+\frac{c}{s}+\mathcal{O}(1)\ .
\end{equation} 
The Hilbert series and TYZ expansion for several examples are given in table \ref{HilbTable}.

\begin{table}
\begin{tabular}{c|c|c}
$L^5$ & Exact Hilbert Series & Small $s$ Expansion\\[5pt] \hline \\[-5pt]
$S^5$ & $\frac{1}{(1-t^{2/3})^3}$ & $\frac{1}{s^3}+\frac{1}{s^2}+\frac{4}{9s}$\\[5pt]
$T^{1,1}$ & $\frac{1+t}{(1-t)^3}$ & $\frac{16}{27 s^3}+\frac{16}{27 s^2}+\frac{8}{27 s}$\\[5pt]
$S^5/\Z_3$ & $\frac{1+7t^2+t^4}{(1-t^2)^3}$ & $\frac{1}{3s^3}+\frac{1}{3s^2}+\frac{4}{27s}$\\[5pt]
$Y^{2,1}$ & $\frac{8t^{2\sqrt{13}/3}f(t)}{27(t-t^{(1+2\sqrt{13})/3})^2(t^5-t^{(1+2\sqrt{13})/3})^2}$ & $\frac{(46 + 13 \sqrt{13})}{324} \frac{1}{s^3} + \frac{(46 + 13 \sqrt{13})}{324} \frac{1}{s^2} + \frac{(22 + 7 \sqrt{13})}{324} \frac{1}{s}$
\end{tabular}
$$f(t)=2t^{22/3}+t^{\frac{2}{3}(4+\sqrt{13})}+3t^{\frac{2}{3}(7+\sqrt{13})}-3t^{\frac{2}{3}(10+\sqrt{13})}-t^{\frac{2}{3}(13+\sqrt{13})}-2t^{\frac{4}{3}(3+\sqrt{13})}$$
\caption{\label{HilbTable}Hilbert Series}
\end{table}

\section{BPS states and free particles}\label{BPS}
In this section we will review how the action for dual giant gravitons can be locally viewed as the action of an electrically charged particle on a K\"ahler-Einstein manifold in the presence of a magnetic field.  This allows us to relate the partition function for dual giant gravitons to the curvature invariants of the K\"ahler-Einstein manifold following the path integral derivation of Douglas and Klevtsov.

We first review the brane construction that leads to the class of theories relevant to this work. Compactifying type IIB string theory to four dimensions on a Calabi-Yau manifold $X^6$ preserves $\mathcal{N}=2$ supersymmetry in 4d. If the Calabi-Yau manifold $X^6$ is a real cone over a Sasaki-Einstein base $L^5$, then placing a stack of $N$ D3-branes at a singular point in $X^6$ leads to a theory with $\mathcal{N} = 1$ supersymmetry on the D3-brane world-volume. The near horizon geometry is $AdS_5\times L^5$, where $L^5$ is a Sasaki-Einstein manifold, called the ``horizon manifold.'' The $AdS$ radial direction $r$ combines with $L^5$ to give a non-compact Calabi-Yau cone $X^6$ with metric 
\begin{equation}
ds_X^2=dr^2+r^2 ds_L^2\ .
\footnote{While we will make extensive use of the assumption that $X^6$ can be written as a cone over a Sasaki-Einstein base, it is important to note that not all non-compact Calabi-Yau manifolds are of this form, and for such exotic spaces that cannot be written as metric cones, it is unclear how to formulate the AdS/CFT correspondence \cite{DRM}.}
\end{equation}
Every Sasaki-Einstein manifold has a distinguished vector field
$$\xi=J(r\frac{\partial}{\partial r})=\frac{\partial}{\partial \psi}$$
called the Reeb vector field.  It is defined using the complex structure $J$ of the Calabi-Yau singularity.
The abelian part of the isometry group of $X^6$ is $(\C^{*})^{s}$ where $s = 1,2$ or 3.  The Reeb vector field can be written as a linear combination of the generators of the $U(1)^s$ isometry group as
$$\xi = \sum_{i} b_i \pp{}{\phi_i}.$$  For certain linear combinations of the $U(1)$ generators, the orbits no longer close and the Sasaki-Einstein manifold is called irregular.  If the orbits close, the Sasaki-Einstein manifold is regular or quasi-regular.

Dual Giant Gravitons are supersymmetric configurations corresponding to D3-branes wrapped on $S^3\times\mathrm{time}\subseteq AdS_5$.  The dynamics of BPS dual giants (the BPS condition fixes the brane's radial position as well as it's position in $B^4$ )  is equivalent to that of BPS point-particles on $L^5$ with Hamiltonian
$$H_{BPS} = \frac{1}{\ell}P_\psi$$
where $\ell$ is the $AdS_5$ radius
$$\ell = 4 \pi g_s N \alpha'^2,$$
and $P_\psi$ is the momentum conjugate to $\psi$.
The phase space for BPS dual giant gravitons is $(X^6, ds^2_X).$  Using geometric quantization, Martelli and Sparks  \cite{Martelli:2006vh} quantize the classical phase space, to obtain the quantum Hamiltonian, $\hat{H}$, for a BPS point-particle.
The partition function dual giant gravitons is
\begin{equation}\label{partitionFn}
Z_{\mathrm{quantum}}(\beta) = \Tr_{\mathcal{H}}e^{-\beta\hat{H}}
\end{equation}
where the trace is over all states in the Hilbert space $\mathcal{H}$.
Martelli and Sparks show that the space of states is precisely the space of holomorphic functions on $X^6$ and that the eigenvalues of the holomorphic functions under the $(\C^{*})^s$ action are ${\bf q}^{{\bf m}} = \prod_{i = 1}^{s} q_i^{m_i}.$ 
They then write the partition function as
\begin{equation}
\Tr_{\mathcal{H}}e^{-\beta\hat{H}} \cong C(q_i = e^{-\beta \xi_i /\ell},X^6)\ .
\end{equation}
where the equivariant character $C(q,X^6)$ is the Hilbert series of $X^6.$

Because the BPS condition fixes the radial position and the position on the K\"ahler-Einstein base, and the Hamiltonian restricted to BPS geodesics given by the Dirac-Born-Infeld and Wess-Zumino terms is precisely that corresponding to a BPS point particle moving in $L^5$ \cite{Martelli:2006vh}. Any Sasakian metric can be written locally in the form
\begin{equation}
ds^2_{L}= ds^2_B+\left(\frac{1}{3}d\psi+A\right)^2
\end{equation}
where $ds^2_B$ is locally a K\"ahler-Einstein metric.
We consider the more general case of geodesic motion on a Lorentzian manifold with metric
\begin{equation}
ds^2=-dt^2+ds^2_{L}=-dt^2+ds^2_B+\left(\frac{1}{3}d\psi+A\right)^2
\end{equation}
where $(L, g_L)$ is a Sasakian manifold, written locally as a $U(1)$ bundle over a base manifold $(B,h_B)$ with Reeb vector $\xi=\partial/\partial\psi$. We may derive geodesic motion from the following action
\begin{equation}\label{5daction}
S=\frac{1}{2}\int d\tau\left[-\dot{t}^2+h_{ij}\dot{x}^i\dot{x}^j+\left(\frac{1}{3}\dot{\psi}+A_i\dot{x}^i\right)^2 \right]\ .
\end{equation}
Absorbing a factor of $3$ in the definition of $p_\psi$, the Hamiltonian is then
\begin{equation}
H=\frac{1}{2}\left[-p_t^2+p_\psi^2+h^{ij}(p_i-p_\psi A_i)(p_j-p_\psi A_j)\right]\ .
\end{equation}
Since $\partial/\partial_t$ and $\partial/\partial_\psi$ are isometries of the system, they must then have conserved conjugate momenta
\begin{equation}
p_t=E,~p_\psi=q\ .
\end{equation}
We look for on-shell solutions, i.e.~those with $H=0$, which tells us the energy is simply
\begin{equation}
E^2=q^2+h^{ij}(p_i-qA_i)(p_j-qA_j)\ .
\end{equation}

Since $h$ is a Euclidean metric, we have the BPS bound
\begin{equation}\label{BPSbound}
E\ge q\ .
\end{equation}
A geodesic which saturates this bound is called a BPS geodesic. The full trajectory of a BPS geodesic is
\begin{equation}
p_i=qA_i,~\dot{x}^i=0,~\dot{\psi}=q\ ,
\end{equation}
and it follows that a BPS geodesic is an orbit of the Reeb vector with fixed momentum $q$. Indeed \cite{Martelli:2006vh} shows an identification between the BPS dual giant graviton phase space and that of the geodesics studied above. This action can be dimensionally reduced to that of a charged particle in an electric field, which is studied by Douglas and Klevtsov in \cite{Douglas:2008pz}. This is motivated by the result from Kaluza-Klein theory \cite{PopeLec}, given a 5d metric with a $U(1)$ isometry,
\begin{equation}
ds^2=h_{ij}dx^idx^j+\phi^2\left(\frac{1}{3}d\psi+A\right)^2\ ,
\end{equation}
$p_\psi$ is constant and the geodesic equation reduces to
\begin{equation}
\ddot{x}^i+\Gamma^i_{jk} \dot{x}^j\dot{x}^k-\frac{q}{m}F^i_j\dot{x}^j=0\ .
\end{equation}
The connection of the bundle is proportional to the base K\"{a}hler form (when $L^5$ is quasi-regular or irregular, this is true away from fixed points), so it is like a magnetic field, $F\sim B_{xy}dx\wedge dy+B_{zw}dz\wedge dw$, as in the work of Douglas and Klevtsov. Indeed, the usual rules of Kaluza-Klein reduction allow us to start with (\ref{5daction}), restrict to BPS states, gauge fix the re-parametrization invariance by choosing $t=\tau$, fix a holomorphic gauge choice for $A$, and add an overall constant shift, which reduces the action to
\begin{equation}\label{SBPS}
S_{\mathrm{BPS},\mathrm{fixed}}=\int dt \left[-1 + h_{i\bar{\j}}\dot{x}^i\dot{\bar{x}}^{\bar{\j}}+A_i\dot{x}^i+\bar{A}_{\bar{\j}}\dot{\bar{x}}^{\bar{\j}}\right]\ ,
\end{equation}
precisely the action Douglas and Klevtsov study in \cite{Douglas:2008pz}.

\section{The proof and a conjecture}\label{Conjecture}

In \cite{Martelli:2006vh}, Martelli and Sparks relate the quantum partition function of free BPS particles moving on the Sasaki-Einstein manifold $L^5$ to the asymptotic number of holomorphic functions on $L^5$.  Simple dimensional analysis tells us that these terms are increasing powers of curvatures. The leading term in the asymptotic expansion of the Hilbert series is determined in terms of the volume of $L^5$. If we consider the entire quadratic polynomial $ak^2+bk+c$, or equivalently all singular terms resulting from the asymptotic expansion of the Hilbert series \ref{Smalls}, we find an interesting result, namely, that these terms can also be expressed in terms of geometric curvature invariants of $L^5$.

The first sub-leading term in the asymptotic expansion of the Hilbert series, $b$, is fixed by the condition that $X^6$ be Calabi-Yau to be twice the leading term $a$; thus, $b$ is also expressible in terms of the volume of $L^5$ \cite{MR0485835}. Alternatively, by the Einstein condition, along with our choice of conventions for the cosmological constant on $L^5$, there is only one independent curvature invariant up to linear order in the Riemann tensor, which we choose to parametrize in terms of the volume. This can also be seen from the CFT data. Stanley \cite{MR0485835} shows that the Hilbert series of a Gorenstein (CY) singularity is ``palindromic'' in the variable t.  A quick exercise shows that when you expand in the variable $s$ after $t = e^{-s}$, this automatically implies that the two leading coefficients are equal.

At the next order in the asymptotic expansion, we find that if our conjecture is to be confirmed, $c$ must be expressible as a linear combination of two independent invariants at quadratic order in the Riemann curvature, which we choose to parametrize by $\mathrm{Riem}^2$ and $R^2$. Furthermore, by fixing the cosmological constant on $L^5$, we may normalize $R=20$, which allows us to express $\int_{L^5}R^2$ in terms of the volume of $L^5$. Thus, there are two independent coefficients which we can fix by matching to $S^5$ and the conifold. 

As a check of the conjecture, we compute the appropriate curvatures for a wide variety of Sasaki-Einstein manifolds with known metrics, including the infinite family $Y^{p,q}$ \cite{Gauntlett:2004yd} as well as several examples from the family $L^{a,b,c}$ \cite{Cvetic:2005vk}, some of which can be found in table \ref{CurvTable}. We then compare to the Hilbert series computed by methods described in \cite{Eager:2010yu} and find the results match in all cases, as is shown in table \ref{HilbTable}.

 \begin{table}
\begin{tabular}{c|c|c}
$L^5$ & Volume & $\int\mathrm{Riem}^2$\\[5pt] \hline\\[-5pt]
$S^5$ & $\pi^3$ & $40\pi^3$\\[5pt] 
$T^{1,1}$ & $\frac{16\pi^3}{27}$ & $\frac{2176\pi^3}{27}$\\[5pt]
$S^5/Z_3$ & $\frac{\pi^3}{3}$ & $\frac{40\pi^3}{3}$\\[5pt]
$Y^{pq}$ & $\frac{\pi^3q^2\left(2+\sqrt{4-\frac{3q^2}{p^2}}\right)}{3p\left(3q^2+p\left(-2p+\sqrt{4p^2-3q^2}\right)\right)}$ & $\frac{8\pi^3\left(2p^2-3q^2+p\sqrt{4p^2-3q^2}\right)\left(106-19\sqrt{4-\frac{3q^2}{p^2}}\right)}{27(p^3-pq^2)}$
\end{tabular}
\caption{\label{CurvTable}Volumes and integrated curvatures}
\end{table}

We prove the conjecture in the restricted case where $L^5$ can be written as a regular $S^1$ bundle over a K\"{a}hler-Einstein manifold $B^4$ by noting that we have related the action studied by Martelli and Sparks in \cite{Martelli:2006vh} to the action studied in \cite{Douglas:2008pz} by Douglas and Klevtsov. Specifically, from the action \ref{SBPS}, Douglas computes the TYZ expansion of the partition function, which we write here for the case of a K\"{a}hler-Einstein manifold
\begin{equation}
\label{DouglasTYZ2} 
g(k)=\frac{1}{2} \left(k^2+k\frac{R}{2}+\left(\frac{1}{24}\mathrm{Riem}^2-\frac{1}{6}\mathrm{Ric}^2+\frac{1}{8}R^2\right)\right)+\mathcal{O}(\frac{1}{k})\ ,
\end{equation}
where we have worked in complex coordinates and followed the standard complex conventions. As explained in equations \ref{Smalls} and \ref{HilbertSum}, the leading behavior of the Hilbert series $H(t)$ is determined by TYZ expansion $g(k)$. We find
\begin{equation}\label{DouglasHilbert}
\lim_{s\rightarrow0}H(e^{-s})=\frac{1}{2} \left( \frac{2}{s^3}+\frac{R/2}{s^2}+ \frac{\left(\frac{1}{24}\mathrm{Riem}^2-\frac{1}{6}\mathrm{Ric}^2+\frac{1}{8}R^2\right)}{s} \right)+\mathcal{O}(1)\ .
\end{equation}
We express these quantities as curvatures on the Sasaki-Einstein space $L^5$, where they are naturally expressed in terms of real curvature conventions\footnote{In order to convert from complex to real curvature conventions, it is useful to note that there are twice as many real coordinates as complex coordinates, and so there is a factor of two from each trace. As an example, $\mathrm{Riem}^2_\C=\frac{1}{16}\mathrm{Riem}^2_\mathbb{R}$.},
\begin{equation}\label{LiftedDims}
\lim_{s\rightarrow0} \pi^3 H(e^{-s})=\frac{\mathrm{vol}}{s^3}+ \frac{3}{2} \frac{\mathrm{vol}}{s^2}+\left(  \frac{3}{2} \right)^2 \left(\frac{91}{216}\mathrm{vol}+\frac{1}{1728}\mathrm{Riem}^2\right)\frac{1}{s}+\mathcal{O}(1) \ .
\end{equation}
Here we have computed the Hilbert series using the degree of the holomorphic functions. The details for the calculation reversing the Kaluza-Klein compactification and lifting the curvatures from $B^4$ to $L^5$ giving equation \ref{LiftedDims} from equation \ref{DouglasHilbert} can be found in appendix \ref{KKlift}. 

Following the work of Martelli, Sparks, and Yau \cite{Martelli:2005tp,Martelli:2006yb} as outlined in section 9 of \cite{Eager:2010yu}, the scaling dimension $\Delta(\O)$ of a mesonic operator in the gauge theory is the degree of the corresponding function. The $\mathcal{N}=1$ superconformal algebra relates the scaling dimensions of chiral primaries to their $R$-charge
\begin{equation}
R(\O)=\frac{2}{3}\Delta(\O)\ .
\end{equation}
This lets us rewrite equation \ref{LiftedDims} as
\begin{equation}\label{Lifted}
\lim_{s\rightarrow0} \left( \frac{2 \pi}{3} \right)^3 H_Q(e^{-s})=\frac{\mathrm{vol}}{s^3}+  \frac{\mathrm{vol}}{s^2}+ \left(\frac{91}{216}\mathrm{vol}+\frac{1}{1728}\mathrm{Riem}^2\right)\frac{1}{s}+\mathcal{O}(1) \ .
\end{equation}

Our proof only applies to the case where $L^5$ is a regular Sasaki-Einstein manifold since it relies on the circle fibers being compact and non-degenerate. Nonetheless, we have shown by example that the conjecture appears to hold even in the quasi-regular case so long as the total space $L^5$ remains smooth. 

\section{Conclusions}\label{Conclusions}
As we have shown, the TYZ asymptotic expansion of the Hilbert series captures significant information about the curvature of the horizon manifold. While we are only able to construct a proof for this fact in the regular case, where the $U(1)$ fibration is compact and non-degenerate, there is significant evidence that our conjecture holds in the quasi-regular case as well. This is a strong indication that there should be an intrinsically 5 dimensional proof of the conjecture, rather than relying on the techniques of Kaluza-Klein reduction. 

Purely from CFT data, we are able to compute $\int_{L^5}\mathrm{Riem}^2$ without reference to an explicit metric, relating $1/N^2$ corrections to the counting of long mesonic operators in the CFT to curvatures in the AdS dual. Such $1/N^2$ corrections, related to the difference in the $a$ and $c$ anomalies of the $\mathcal{N}=1$ SCFT, have recently been shown to be closely related to the appearance of higher order curvature corrections in the dual gravitational action \cite{Liu:2010gz}. Thus, it should come as no surprise that the AdS/CFT correspondence allows us to extract curvature information from such corrections. Nonetheless, given the relative scarcity of examples of the emergent geometry, we should attempt to fully exploit any novel constructions such as this to increase our knowledge. Hopefully, through careful study, it will be possible to generalize this result and relate other details of the CFT to the emergence of bulk geometry.

There are a large number of open questions and future directions to look into. The most obvious one is extending our proof to the quasi-regular and irregular cases. It should be possible to prove a generalization of  (\ref{DouglasTYZ2}) with proper treatment of the singular points of the fiber. It would also be interesting to know if there is a directly five dimensional method of calculating the TYZ expansion, in which case we would not have to worry about any of the issues in the reduction to the K\"ahler-Einstein base.

Another question is what further geometric information is hidden in the full matrix $H(Q;t)$. The full matrix appears to encode information about the baryonic spectrum of the gauge theory \cite{Berenstein:2002ke} and it would be interesting to relate this to the emergent geometry.

A third question is how such heat kernel expansions carry over to compactifications using generalized geometry. Given that it was recently shown that the volume-minimization/a-maximization story translates \cite{Gabella:2010cy}, it would be interesting to study what generalized curvature invariants appear in subleading terms in the heat kernel expansion.

Lastly, we have studied $1/N^2$ corrections to the CFT which are hiding nontrivial geometric information about the gravity side. Is there geometric information hidden in $1/\lambda$ corrections as well? Given that such $\alpha'$ corrections to the supergravity theory include higher curvature corrections, these must be related. It would be very interesting to connect these to the related work in \cite{Liu:2010gz}.

\vskip 1cm
\centerline{\bf Acknowledgements}
\vskip 0.5cm
It is a pleasure to thank D. Berenstein, M.R. Douglas, and D.R. Morrison for helpful discussions.
The work of RE was supported in part by the National Science Foundation under grants DMS-0606578 and DMS-1007414 and by the World Premier International Research Center Initiative (WPI Initiative), MEXT, Japan.  RE would like to thank IPMU, where part of this work was completed, for their hospitality.
The work of MG was supported in part by the U.S. Dept. of Energy under Contract DE-FG02-91ER40618.
The work of MMR was supported in part by NSF grant PHY-0855415 and the Simons Postdoctoral Fellowship Program. Much of this work was completed while MMR was at UCSB.

\appendices

\section{Kaluza-Klein lift} \label{KKlift}

The near horizon geometry is given by $AdS_5 \times L^5$. To obtain a total space $X^6$ which is Calabi-Yau, we require the base space $L^5$ to be Einstein \cite{Morrison:1998cs,Acharya:1998db}, e.g.
\begin{equation}
R_{ab}=4 g_{ab}\ ,
\end{equation}
where we fix the 5d Einstein coefficient to be 4 by convention (we will use $a,b,\ldots$ to denote indices along $L^5$). 

We also know that such an Einstein manifold $L^5$ must have a $U(1)$ isometry and therefore is Sasakian. Thus $L^5$ may always be written as a principal $S^1$ bundle $\pi: L^5\rightarrow B^4$ over a K\"ahler-Einstein base with a connection 1-form $\eta$ \cite{Martelli:2005tp,Martelli:2006yb}. The bundle structure allows the metric on $L^5$ to be locally written in the form
\begin{equation}
\five{g}=\pi^*\left(\four{h}\right) + \phi^2\eta\otimes\eta\ .
\end{equation}
Making contact with the language of Kaluza-Klein compactifications, $\phi$ plays the role of the dilaton field, while the connection 1-form $\eta = \frac{1}{3}d\psi + A$, where $A$ is a $U(1)$ gauge field on the base $B$ and $\psi$ is a coordinate on the circle fiber. In order for both $B$ and $M$ to satisfy the Einstein condition, $\phi$ must satisfy the equation $\four{\Box}\phi=\mathrm{const}$. However, the only smooth solutions to this equation for compact $B$ are constant $\phi$, and thus we may henceforth restrict ourselves to the case of {\em constant} dilaton field $\phi=1/2$. 

Since the structure group $U(1)$ is abelian,
\begin{equation}
d \eta = \Gamma
\end{equation}
where $\Gamma$ is the curvature form of $\eta$.  In physicist's terms
\begin{equation}
d \eta = \frac{1}{3}d^2 \psi + d A = dA
\end{equation}
so we can write
\begin{equation}
\Gamma = \pi^{*}\left( \sum_{i,j} F_{ij} \theta^i \wedge \theta^j \right)
\end{equation}
where $F_{ij} = - F_{ji}$ is the curvature of the $U(1)$ bundle on the base and $\theta^i$ form an orthonormal basis of 1-forms on $B$. (Above and throughout what follows, we will use roman indices $i,j,\ldots$ to denote directions on the base and $\psi$ to denote directions along the circle fiber.) Furthermore, it will be essential for the computations that follow that we work in an orthonormal frame, since then on the bundle, covariant derivatives in bundle directions are equivalent to covariant derivatives on the base. 

\begin{prop}
The Riemann curvature tensor on $L$ can be expressed as the curvature tensor of the base $B$ and the curvature of the $U(1)$ connections as
\begin{align}
\five{R}_{ijkl} & = \four{R}_{ijkl} - \phi^2 \left( 2 F_{ij}F_{kl} + F_{ik} F_{jl} - F_{il}F_{jk} \right) \label{ijkl}\\
\five{R}_{i\psi k\psi} & = \phi^2 \left( \sum_{l} F_{il} F_{kl} \right) \label{ipsikpsi}\\ 
\five{R}_{i\psi kl} & = \phi \left( \nabla_{l} F_{ik} - \nabla_k F_{il} \right) = -\phi \nabla_{i} F_{kl} \label{ipsikl}
\end{align}
The identity
\begin{equation}
\nabla_{l} F_{ik} - \nabla_k F_{il}  = - \nabla_{i} F_{kl}
\end{equation}
follows from $\sum_{ij} F_{ij} \theta^i \wedge \theta^j$ being closed.
\end{prop}

\begin{prop}
If $F_{ij} = 2 J_{ij}$ then
\begin{align}
\five{R}_{ij} & = \four{R}_{ij} - 2 h_{ij} \\
\five{R}_{i\psi} & = 0\\
\five{R}_{\psi\psi} & = n
\end{align}
\end{prop}

We need to relate the Riemann curvature tensor on $L^5$ to the Riemann curvature tensor and Maxwell stress tensor in the Kaluza-Klein reduction to $B^4$. By the Einstein condition, at second order in curvature there are only two independent quantities, which we choose to express in terms of the volume and $\mathrm{Riem}^2$. By our normalization convention for $\phi$, which fixes $\psi$ to have period $2\pi$, $\five{\mathrm{Vol}} = 2\pi\four{\mathrm{Vol}}$. Furthermore, since the Maxwell tensor is proportional to the complex structure on $B$, we will be able to express $\five{\mathrm{Riem}^2}$ purely in terms of $\four{\mathrm{Riem}^2}$ and $\four{\mathrm{Vol}}$. 

Before we continue with the computation, it is worth reminding ourselves of a few useful facts. First, recall the symmetries of the Riemann tensor, namely
\begin{equation}
R_{ijkl}=-R_{ijlk}=-R_{jikl}=R_{klij}
\end{equation}
along with the Bianchi identity
\begin{equation}
R_{ijkl}+R_{iklj}+R_{iljk}=0\ .
\end{equation}
Given our convention that $\four{R}_{ij}=6h_{ij}$, $\four{R}=24$. 
Also recall that the complex structure $J_{ij}$ is antisymmetric and $J^2=-1$. 
Furthermore, it is worth noting that, in the case we are considering 
\begin{equation}
\five{R}_{i\psi kl} = -\phi \nabla_i F_{kl} = -2\phi \nabla_i J_{kl} = 0\ ,
\end{equation}
since the condition $\nabla_i J_{kl} = 0$ is simply the integrability condition for the complex structure on the base, making the base a K\"ahler manifold. 
Finally, the compatibility of the complex structure with the covariant derivative on the base also allows us to derive one final identity, 
\begin{eqnarray*}
\four{R}_{ijkl}J^{il}J^{jk} &=& h(\four{R}(\partial_k,\partial_l) \partial_i,\partial_j)  J^{il} J^{jk} \\
&=& h(J^{il} \four{R}(\partial_k, \partial_l) \partial_i, J^{jk} \partial_j) \\
&=& h( \four{R}(\partial_k, \partial_l) J^{il} \partial_i, J^{jk} \partial_j) \\
&=& h( \four{R}(\partial_k, \partial_l)  \partial_i,  \partial_j) \\
&=& - \four{R}\ .
\end{eqnarray*}

Armed with these facts, we are now prepared to proceed with our computation. 
\begin{eqnarray*}
\five{\mathrm{Riem}^2} &=& \five{R}_{ijkl}\five{R}^{ijkl} + 4\left(\five{R}_{i\psi k\psi}\five{R}^{i\psi k\psi}\right)\\
&=& \left(\four{R}_{ijkl} - \phi^2 \left( 2 F_{ij}F_{kl} + F_{ik} F_{jl} - F_{il}F_{jk} \right)\right)^2 + 4\phi^2F_{ik}F^{ik}\\
&=&\begin{array}{ll}\four{\mathrm{Riem}^2} - 8\phi^2\four{R}_{ijkl}\left(2J_{ij}J_{kl} + J_{ik}J_{jl} - J_{il}J_{jk} \right)\\
+16\phi^4\left(2J_{ij}J_{kl} + J_{ik}J_{jl} - J_{il}J_{jk} \right)^2 + 16\phi^2J_{ik}J^{ik}\end{array}\\
&=&\four{\mathrm{Riem}^2} - 2(6\ \; \four{R}) + 120 + 16\\
&=&\four{\mathrm{Riem}^2} - 152\ .
\end{eqnarray*}

While this equality is purely geometric and holds locally, our conjecture arises from the CFT dual and depends on integrated quantities. While the horizon Calabi-Yau manifolds $X^6$, and hence the Sasaki-Einstein 5-folds $L^5$, are generically not homogeneous spaces, our count of the number of dual giant gravitons is global (i.e. integrated) and thus does not probe this inhomogeneity. Therefore, we will make use of an integrated version of this relation, 
\begin{equation}\label{IntRiemConjecture}
\frac{\int_{L^5}\five{\mathrm{Riem}^2}}{\mathrm{Vol}(L^5)} = \frac{\int_{B^4}\four{\mathrm{Riem}^2}}{\mathrm{Vol}(B^4)} - 152\ .
\end{equation}
  


\end{document}